\documentstyle[11pt,epsfig,subfigure]{article}
\newcommand{\beq}{\begin{equation}}
\newcommand{\eeq}{\end{equation}}
\newcommand{\ba}{\begin{array}}
\newcommand{\ea}{\end{array}}
\newcommand{\EQN}{\label}
\newcommand{\dsp}{\displaystyle}
\newcommand{\api}{\frac{\alpha_s}{\pi}}
\newcommand{\as}{\alpha_s}
\newcommand{\msbar}{\overline{\mbox{MS}}}
\newcommand{\ordas}{{\cal O}(\alpha_s}

\begin{document}
\begin{titlepage}
\noindent
%
%
\mbox{}
\hfill LBL--37269 \\
\mbox{}
\hfill TTP 95--23 \\
\mbox{}
\hfill May 1995

\protect\vspace*{.8cm}
%
%
%
\begin{center}
  \begin{Large}
  \begin{bf}
Second Order QCD
 Corrections to  Scalar and Pseudoscalar
 Higgs Decays into Massive Bottom Quarks$^{\dagger}$
   \\
  \end{bf}
  \end{Large}
%
%
  \vspace{0.3cm}
K.G.~Chetyrkin$^{ab \sharp}$,
A.~Kwiatkowski$^{c \star}$
\\
%

\begin{itemize}
\item[$^a$]
 Institute for Nuclear Research,
 Russian Academy of Sciences   \\
 60th October Anniversary Prospect 7a,
 Moscow 117312, Russia
\item[$^b$]
    Institut f\"ur Theoretische Teilchenphysik,
    Universit\"at Karlsruhe \\
    D-76128 Karlsruhe, Germany
\item[$^c$]
              Theoretical Physics Group,
              Lawrence Berkeley Laboratory\\
              University of California,
              Berkeley, CA. 94720, USA
\end{itemize}
\vspace{0.2cm}

%
  \vspace{0.5cm}
  {\bf Abstract}
\end{center}
\begin{quotation}
\noindent
Quark mass effects in ${\cal O}(\alpha_s^2)$
 QCD corrections to the
decay rates of intermediate Higgs bosons
 are studied.
The total hadronic rate and the
 partial decay rate
 into bottom quarks
are analyzed for  the Standard (scalar) Higgs boson
as well as for   pseudoscalar Higgs bosons.
The calculations of three different contributions
are presented.
First,
the flavour singlet diagrams containing
two closed fermion loops are calculated for a
nonvanishing bottom mass
in the heavy top limit.
Their leading contribution, which is of the
same order as the quasi-massless
nonsinglet corrections,
and the subleading terms are found.
Large logarithms arise due to
the separation of the pure gluon final state
from the  bottom final states.
Second,
quadratic bottom mass corrections originating from
nonsinglet diagrams are presented.
Third,
nonsinglet corrections
induced by  virtual heavy top quarks
are calculated  in leading and subleading orders.
It is demonstrated that,
in order $\alpha_s^2$,  the first  contribution
numerically dominates over the second and the
third ones.

\end{quotation}

\vfill

\footnoterule
\noindent
$^{\dagger}${\footnotesize This work was in part supported by
                           US DoE under Contract DE-AC03-76SF00098.}

\noindent
$^{\sharp}${\footnotesize Supported by Deutsche Forschungsgemeinschaft,
                          grant no. Ku 502/6-1.}

\noindent
$^{\star}${\footnotesize Supported by Deutsche Forschungsgemeinschaft,
                          grant no. Kw 8/1-1.}

\end{titlepage}

\renewcommand{\thepage}{\roman{page}}
\setcounter{page}{2}
\mbox{ }

\vskip 1in

\begin{center}
{\bf Disclaimer}
\end{center}

\vskip .2in

\begin{scriptsize}
\begin{quotation}
This document was prepared as an account of work sponsored by the United
States Government. While this document is believed to contain correct
 information, neither the United States Government nor any agency
thereof, nor The Regents of the University of California, nor any of their
employees, makes any warranty, express or implied, or assumes any legal
liability or responsibility for the accuracy, completeness, or usefulness
of any information, apparatus, product, or process disclosed, or represents
that its use would not infringe privately owned rights.  Reference herein
to any specific commercial products process, or service by its trade name,
trademark, manufacturer, or otherwise, does not necessarily constitute or
imply its endorsement, recommendation, or favoring by the United States
Government or any agency thereof, or The Regents of the University of
California.  The views and opinions of authors expressed herein do not
necessarily state or reflect those of the United States Government or any
agency thereof, or The Regents of the University of California.
\end{quotation}
\end{scriptsize}

\vskip 2in

\begin{center}
\begin{small}
{\it Lawrence Berkeley Laboratory is an equal opportunity employer.}
\end{small}
\end{center}

\newpage
\renewcommand{\thepage}{\arabic{page}}
\setcounter{page}{1}

\section{Introduction}
\renewcommand{\arraystretch}{2}
In the last years the Standard Model (SM) has
faced a remarkable lot of
experimental tests, which were
 performed at LEP and SLC with
very high precision. Due to the overwhelming
 agreement between
theory and experiments the description
 of the world of elementary
particles through a $SU(3)_C\times SU(2)_L
\times U(1)_Y$ gauge theory
has emerged as a profoundly tested and
firmly established theoretical framework.
Despite of these achievements the detailed nature
of electroweak symmetry breaking is still
waiting for experimental confirmation. The search
for a physical Higgs boson and the study of
its properties will be the main subject of future
collider experiments.

 Standard Model
 properties
of the Higgs boson have been discussed
in many reviews (see for example
\cite{GunHabKanDaw90,Kni93}).
In the minimal SM one physical scalar
 Higgs boson is present as a remnant of
the mechanism of mass generation.
  Particularly interesting for the observation
 of the Higgs boson with  an
  intermediate mass
$M_H<2M_W$ is the dominant decay channel
into a bottom pair
$H\rightarrow b\bar{b}$.
The  partial width
$\Gamma(H\rightarrow b\bar{b})$ is
significantly affected by QCD
radiative  corrections.
First order $\ordas)$ corrections
including the full $m_b$ dependence
 were studied by several groups
\cite{BraLev80,Sak80,InaKub81,DreHik90,SurTka90,KatKim92}.
Second order corrections were calculated in the
limit $m_b^2 \ll M_H^2$. Apart from the trivial
overall factor $m_b^2$ due to the Yukawa coupling,
corrections were obtained for otherwise massless
quarks
by \cite{GorKatLar84,GorKatLarSur90} and for
a nonvanishing  mass of the virtual top quark
 by \cite{Kni94}.
Subleading quadratic mass corrections in the
$m_b^2/M_H^2$ expansion were found in
\cite{Sur94a}.
In this work we complete the discussion
of quark mass effects in second order
QCD corrections.
Our calculations provide the so far missing
contributions and in part also serve
as a  crosscheck for existing results.

Theories beyond the SM are usually characterized
by an enlarged Higgs sector and may allow for
different quantum numbers of the Higgs bosons.
For example, one of the most appealing extensions
of the SM,
the Minimal Supersymmetric Standard Model (MSSM),
contains
two complex isodoubletts with opposite
hypercharge (see e.g. \cite{GunHabKanDaw90}),
resulting in
five mass eigenstates of the Higgs fields:
 two scalar (CP-even) neutral
$H^0,h^0$, one pseudoscalar
(CP-odd) neutral $A$ and two
charged $H^{\pm}$ physical Higgs bosons.
As a consequence QCD corrections to the fermionic
decays of a pseudoscalar Higgs
have been studied in the past in many works
\cite{BecNarRafYnd81,Bro81,SurTka90,Sur94b,DjoGam94}.
This was motivation enough to carry out our
analysis of quark mass effects in decay rates
 of pseudoscalar
Higgs bosons as well.
 Our formulae are tailored in such a way that they are
 immediately applicable to the MSSM.
However, the commitment towards this specific
choice of model  is limited by the fact
that supersymmetric QCD with the exchange of gluinos
is is not covered in this work.

Let us start our analysis  of QCD corrections
to the decay rates of the neutral Higgs bosons
into bottom quarks  with
the two point correlators
\beq \EQN{i1}
\Pi^{S/P}(q)
=i\int dx e^{iqx}\langle 0|\;T\;J^{S/P}(x)
J^{S/P\dagger}(0)\;|0\rangle
\eeq
of the scalar current
 $J^S=\bar{\Psi}_f\Psi_f$
and the pseudoscalar current
 $J^P=\bar{\Psi}_f i\gamma_5\Psi_f$
for quarks with flavour $f$
and  mass $m_f$, which are coupled to the scalar
Higgs bosons (we use the generic notation
$H$ for $H^0$ and $h^0$) and the pseudoscalar
one ($A$) respectively.

The analysis of quark mass effects in second order
QCD corrections is the main concern of this work.
However,
 the calculation of
$\ordas^2)$ corrections to the
 correlators keeping the
exact quark mass dependence
would be an enormous task.
Fortunately only the first few terms
of the expansion in the small parameter $m_b^2/s$
represent already a very good approximation
for processes at high energies.
The partial decay rates of the Higgs bosons
into bottom quarks
can therefore be written in the form
 \beq \EQN{i3}
\Gamma_{b\bar{b}}
=\frac{3G_F}{4\sqrt{2}\pi}M_H \bar{m}_b^2
C^{S/P}_{bb} R^{S/P}
\eeq
where all information is contained in the
absorptive part of the corresponding
current correlator:
\beq \EQN{i4}
\ba{ll}
\dsp
R(s)
& \dsp
 = \frac{8\pi}{3s}
\mbox{\rm Im} \Pi(s+i\epsilon) \\
& \dsp
= 1 + \Delta\Gamma_1 \left(\api\right) +
 \left(\api\right)^2
 \left[\Delta\Gamma_2
       +\frac{s}{m_t^2}
       \tilde{\Delta\Gamma}_2\right]
\\ & \dsp
+\frac{\bar{m}_b^2}{s}\left(
\Delta\Gamma_0^{(m)}
+ \Delta\Gamma_1^{(m)} \left(\api\right)
  +  \left(\api\right)^2
\left[\Delta\Gamma_2^{(m)}
     +\frac{s}{m_t^2}\tilde{\Delta\Gamma}_2^{(m)}
\right]
\right)
\\ & \dsp
+{\cal O}\left(\frac{\bar{m}_b^4}{s^2}\right)
+{\cal O}\left(\frac{s^2}{m_t^4}\right)
+{\cal O}\left(\frac{\bar{m}_b^2 s}{m_t^4}\right)
{}.
\ea\eeq

\begin {figure}
\begin{center}
\begin {tabular}{cc}
\subfigure[]{\epsfig{file=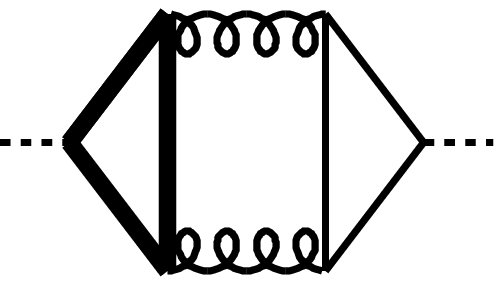,width=5.cm,height=5.cm}}
&
\subfigure[]{\epsfig{file=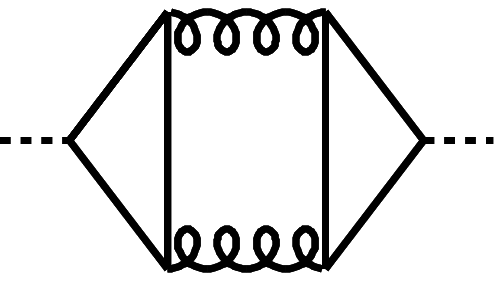,width=5.cm,height=5.cm}}
\end {tabular}
\end{center}
\caption {
Diagrams of the $\ordas^2)$ singlet contribution. (Thick lines: top,
thin lines: bottom, curly lines: gluon, dashed lines: Higgs)
}
\end{figure}

\begin {figure}
\begin{center}
\begin {tabular}{cc}
\subfigure[]{\epsfig{file=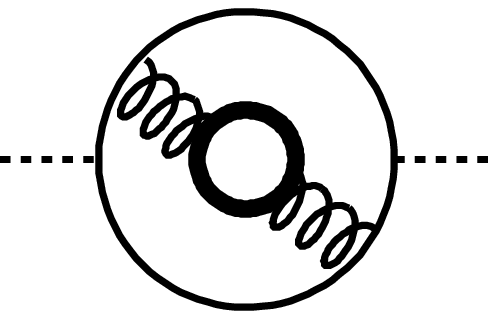,width=5.cm,height=5.cm}}
&
\subfigure[]{\epsfig{file=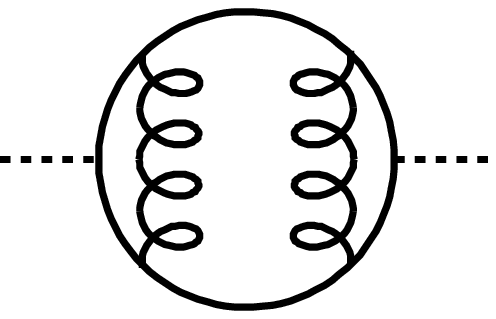,width=5.cm,height=5.cm}}
\end {tabular}
\end{center}
\caption {
Examples of $\ordas^2)$ nonsinglet diagrams.
}
\end{figure}

The coefficients $C^{S/P}_{ff'}=
g^{S/P}_fg^{S/P}_{f'}$ are the respective
weights of the various couplings  between
Higgs bosons and fermions.
For the MSSM  they are given
 by
$ g^{H^0}_{up}= \sin\alpha/\sin\beta,\;
 g^{H^0}_{down}= \cos\alpha/\cos\beta,\;
 g^{h^0}_{up}= \cos\alpha/\sin\beta,\;
 g^{h^0}_{down}= -\sin\alpha/\cos\beta,\;
 g^{A}_{up}= \cot\beta$ and
 $g^{A}_{down}= \tan\beta$
with $\alpha,\beta$ denoting the usual
mixing angles (see e.g. \cite{GunHabKanDaw90}).
All formulae in this work
are
 applicable for the Higgs decay
in the mimimal SM with $g^S_f=1,g^P_f=0$.

The outline of this work is as follows.
The decay rate of the  Higgs
boson $A$ involves  traces over fermion
loops which are separately coupled to
a pseudoscalar current. For this case
some technical details  about
the treatment of $\gamma_5$ in $D=4-2\epsilon$
dimensions are given in Section 2.
Flavour singlet contributions  to the partial
decay rates are discussed in Section 3.
The heavy triangle diagram of Figure 1a
contains two closed
fermion loops composed
of a top quark and  a bottom quark respectively.
It gives rise to a contribution to
$\Delta\Gamma_2$, which
is of the same order
as the otherwise massless nonsinglet corrections.
For the scalar Higgs boson  it
was
 presented in \cite{CheKueKwi94}
some time ago.
Subleading terms  $\tilde{\Delta\Gamma}_2$ are
also calculated in Section 3.
The light triangle
graph of Figure 1b consisting only of bottom
loops is suppressed by $m_b^2/s$.
Its contribution to $\Delta\Gamma_2^{(m)}$,
obtained  by
\cite{Sur94a} for the scalar Higgs, is only part
of the complete answer, since bottom mass corrections
of the same order originate also from the
heavy triangle diagram and have to be added.
This is done for the scalar and the pseudoscalar
case in Section 3. Again subleading terms
$\tilde{\Delta\Gamma}_2^{(m)}$ are given.

In Section 4 nonsinglet contributions (see
Figure 2) to
the partial Higgs decay rates are discussed.
Quadratic mass corrections to $R^{S/P}$
in second order are  presented in this section.
Their contributions to $\Delta\Gamma_2^{(m)}$
are obtained
from a recent calculation of the corresponding
current correlators \cite{Che95}
and agree with \cite{Sur94a,Sur94b}.
  Furthermore the
calculation of the double bubble diagram of
Figure 2a containing a virtual top quark loop
is performed in the heavy top limit.
For vanishing bottom mass the exact top dependence
 of this contribution was obtained in \cite{Kni94}.
Our result for
$\tilde{\Delta\Gamma}_2$ agrees with the leading
(power-suppressed) term in the $M_H^2/m_t^2$
expansion of \cite{Kni94}.
In addition we present the correction for
non-vanishing bottom masses, thus completing
the second order corrections
$\Delta\Gamma_2,
\tilde{\Delta\Gamma}_2,
\Delta\Gamma_2^{(m)},
\tilde{\Delta\Gamma}_2^{(m)}$.

In Section 5 the results obtained in the full six flavour
theory are expressed in the framework of an effective theory
 with five active quark flavours.

The  numerical size of
the corrections and conclusions are
given in Section 6.

\section{The Treatment of $\gamma_5$}
The calculations of the two-point correlators
are performed with dimensional regularization
in the $\msbar$-scheme. For the pseudoscalar
correlator  we employ the definition of $\gamma_5$
in $D\neq 4$ dimensions
as suggested  by 't Hooft
and Veltman \cite{HooVel72}.

Following \cite{Lar92} we avoid an explicit separation
of Lorentz indices into 4 and ($D$-4) dimensions
 and   work,  in $D=4-2\epsilon$ dimensions,
with
the generalized current
\beq\EQN{c1}
P^{[\rho\mu\nu\lambda]} =
\bar{\Psi} i\gamma^{[\rho\mu\nu\lambda]} \Psi
\eeq
where
$
 \gamma^{[\rho\mu\nu\lambda]}=
(\gamma^{\rho}\gamma^{\mu}
 \gamma^{\nu}\gamma^{\lambda}
+\gamma^{\lambda}\gamma^{\nu}
 \gamma^{\mu}\gamma^{\rho}
-\gamma^{\mu}\gamma^{\nu}
 \gamma^{\lambda}\gamma^{\rho}
-\gamma^{\rho}\gamma^{\lambda}
 \gamma^{\nu}\gamma^{\mu})/4
$.
Taking the limit $D\rightarrow 4$ at the end of the
calculation, when the result is finite,
the pseudoscalar current is recovered by
\beq\EQN{c2}
j^P = \frac{i}{4!}\epsilon_{\rho\mu\nu\lambda}
P^{[\rho\mu\nu\lambda]}
.\eeq
The corresponding generalized current
correlator has the following form
\beq\EQN{c3}
\ba{ll}\dsp
\Pi^{[\rho\mu\nu\lambda]}_{[\rho'\mu'\nu'\lambda']}
(q^2)
&\dsp
=i\int dx e^{iqx}\langle 0|\;T\;
P^{[\rho\mu\nu\lambda]}(x)
P^{[\rho'\mu'\nu'\lambda']}(0)\;|0\rangle
\\ & \dsp
={[\rho\mu\nu\lambda] \atop
  [\rho'\mu'\nu'\lambda']}
\Pi_1(q^2)
\ea
\eeq
where
${[\rho\mu\nu\lambda] \atop
  [\rho'\mu'\nu'\lambda']} = (1/4!){\rm det}
(g_{\alpha\alpha'})$
with
$\alpha=\rho,\mu,\nu,\lambda$
and
$\alpha'=\rho',\mu',\nu',\lambda'$.
It is convenient to work with the
contracted tensor
$\Pi^{[\rho\mu\nu\lambda]}_{[\rho\mu\nu\lambda]}$, which
is related by
\beq\EQN{c4}
\Pi^P = \frac{1}{4!} \left(
1+8\epsilon+52\epsilon^2+\frac{976}{3}
\epsilon^2\right)
\Pi^{[\rho\mu\nu\lambda]}_{[\rho\mu\nu\lambda]}
\stackrel{D\rightarrow 4}{\longrightarrow}
\frac{1}{24}
\Pi^{[\rho\mu\nu\lambda]}_{[\rho\mu\nu\lambda]}
\eeq
to the pseudoscalar current correlator
$\Pi^P$.

\section{Singlet Contributions}
The flavour singlet diagrams are characterized
 by two closed quark loops which are connected by
gluons only. The heavy triangle graph of
Figure 1a containing a top as well as a bottom loop
is calculated in the heavy top mass limit. In
leading order it contributes to the quasi-massless
corrections $\Delta\Gamma_2$: The Yukawa couplings
of the fermions yield a factor $m_bm_t$. In addition
each fermion trace produces a factor $m_b$ and
$m_t$ respectively. Together with a power
suppression of $1/m_t^2$ on dimensional grounds
all mass factors are combined to $m_b^2$ and
are thus of the same order as the otherwise
massless nonsinglet corrections.

This leading and the subleading power suppressed
$s/m_t^2$ contributions are obtained by employing
the  hard mass procedure
\cite{PivTka84,GorLar87,CheSmi87,Smi91}, which
effectively is an expansion in the
 inverse heavy mass $1/m_t^2$ and has
sucessfully been used
in previous works \cite{CheKwi92,CheKwi93,CheTar93}
 for the calculation of
singlet diagrams for the decay rate of the
$Z$ boson.
All possible ``hard'' subgraphs containing the
heavy particle are selected and
expanded with
 respect to the external momenta and
the small masses.
This formal Taylor expansion
 is inserted as an
effective vertex.

Having integrated out the heavy top
 $m_t^2\gg M_H^2,m_b^2$ the remaining diagram still
 contains two different scales
 $M_H^2\gg m_b^2$, where the Higgs mass
comes into play through the external
momentum of the propagator integral.
In analogy to the hard mass procedure
the expansion
in $m_b^2/q^2$ is obtained through the hard
momentum procedure, where all possible  subgraphs,
through which the heavy momentum $q$ may be
routed,
are reduced to a dot and expanded with
respect to  $m_b$ and eventual small momenta
as compared
to $q$. In this way we have calculated the
leading and next-to-leading order of the
$m_b^2/s$ series.
As a result of the  expansions
the computation is simplified
due to a factorization of the integrals.
 The three loop diagram  decomposes
into the product of massive tadpole integrals
 and massless propagator integrals.
The latter are computed with the help
 of the multiloop program MINCER
\cite{LarTkaVer91} written in the
 symbolic manipulation language FORM
\cite{Ver91}.

For the light triangle diagram of Figure 2b
containing two bottom loops
we also perform an expansion in $m_b^2/s$.
As already stated in \cite{Sur94a}, the first
nonvanishing contribution is suppressed by
$m_b^2/s$ and reproduced by our calculation.
It has to be combined with the corrections of
the same order originating from the
heavy triangle diagram. Additionally we find
 that the corresponding contribution from the
pseudoscalar light triangle graph vanishes
 identically.

The absorptive parts
of the singlet
diagrams (Figures 1a, 1b)
 are given by
\beq \EQN{s1}
\ba{rl}\dsp
\Delta\Gamma^{triangle}
(H\rightarrow b\bar{b},{\rm gluons})
& \dsp
= \frac{3G_F}{4\sqrt{2}\pi} M_H \bar{m}_b^2
\left(\api\right)^2
\\ \dsp
\cdot \Bigg\{ C^S_{tb}\Bigg(
& \dsp
\frac{28}{9}-\frac{2}{3}\ln
 \frac{M_H^2}{m_t^2}
\\ & \dsp
+ \frac{M_H^2}{m_t^2}
\left[\frac{2011}{24300} - \frac{41}{1620}
\ln\frac{M_H^2}{m_t^2}\right]
\\ & \dsp
+ \frac{\bar{m}_b^2}{M_H^2}
\left[ -10+4\ln\frac{M_H^2}{m_t^2}
      +\frac{4}{3}\ln\frac{\bar{m}_b^2}{M_H^2}
\right]
\\ & \dsp
+ \frac{\bar{m}_b^2}{m_t^2}
\left[ \frac{713}{2700}
      -\frac{7}{270}\ln\frac{M_H^2}{m_t^2}
      -\frac{1}{54}\ln\frac{\bar{m}_b^2}{M_H^2}
\right]\Bigg)
\\
\dsp +C^S_{bb} & \dsp
\frac{\bar{m}_b^2}{M_H^2}8
\Bigg\}
\ea
\eeq

\beq \EQN{s1a}
\ba{rl}\dsp
\Delta\Gamma^{triangle}
(A\rightarrow b\bar{b},{\rm gluons})
& \dsp
= \frac{3G_F}{4\sqrt{2}\pi} M_A \bar{m}_b^2 C^P_{tb}
\left(\api\right)^2
\\ \dsp
\cdot \Bigg\{
& \dsp
4-\ln
 \frac{M_A^2}{m_t^2}
\\ & \dsp
+ \frac{M_A^2}{m_t^2}
\left[\frac{61}{324} - \frac{7}{108}
\ln\frac{M_A^2}{m_t^2}\right]
\\ & \dsp
+ \frac{\bar{m}_b^2}{M_A^2}
\left[ -5+2\ln\frac{M_A^2}{m_t^2}
      -2\ln\frac{\bar{m}_b^2}{M_A^2}
\right]
\\ & \dsp
+ \frac{\bar{m}_b^2}{m_t^2}
\left[ \frac{19}{108}
      -\frac{1}{18}\ln\frac{M_A^2}{m_t^2}
      -\frac{5}{18}\ln\frac{\bar{m}_b^2}{M_A^2}
\right]\Bigg\}
\ea
\eeq
They are constituted by the sum of all possible
cuts of the diagrams and represent the decays
into two ($b\bar{b},gg$), three ($b\bar{b}g$)
and four particle ($b\bar{b}b\bar{b}$) final
states.

Note that the expressions (8) and (9)
contain no explicit $\mu$ dependence.
This is  a direct consequence
 of the fact that the singlet diagrams contain no divergent
subgraphs. This, in turn, means that their absorptive parts
and, thus, (8), (9) have no dependence on the choice of the
renormalization scheme.

The contributions of
 additional ``ultralight'' singlet diagrams
containing
a bottom and another light ($f=u,d,s,c$) quark
loop deserve some extra discussion with respect
to their assignment to partial decay rates
into specific quark species.
 The absorptive parts of these diagrams comprise
again
 two ($b\bar{b},f\bar{f},gg$), three
($b\bar{b}g,f\bar{f}g$)
and four particle ($b\bar{b}f\bar{f}$) final
states. The cuts with
 two fermion final states
($b\bar{b}(g),f\bar{f}(g)$) should be
calculated separately and could uniquely be
assigned to the partial rates
$\Gamma_{b\bar{b}}$ and $\Gamma_{f\bar{f}}$
respectively.
The situation is different for the four
fermion final state ($b\bar{b}f\bar{f}$).
An unambiguous assignment to a specific partial
rate is not possible in this case. The question
to which partial rate this piece should
most reasonably  be
counted must be decided according to the
peculiarities of the experimental setup and
identification methods.
However, since the ultralight singlet contributions
are proportional to $m_f^2m_b^2$, they are much
 smaller than the already small $m_b^4$
contribution from
the double bottom triangle graph.
We therefore have adopted
a pragmatic point of view and neglected the
 ul\-tra\-light terms in eq. (\ref{s1}).
(For the decay of the pseudoscalar Higgs they
vanish identically.)
No  complications of this kind
arise for the total Higgs
decay into hadrons. We agree to the result in
\cite{Sur94a}  for the light and ultralight
singlet contributions to $\Gamma_{{\rm had}}$.
However, in view of the above discussion we find
that in this work the
assignment of the
 triangle graphs to the
 partial rates $\Gamma_{f\bar{f}}$
 is not well motivated.
The complete formula for
$\Gamma_{{\rm had}}$ will be given
in Section 5.

In order to complete our calculation and to
  arrive at the
singlet result
for the decay rate $\Gamma_{b\bar{b}}$
into bottom quarks
one still  needs to subtract
from eqs. (\ref{s1}),(\ref{s1a}) the
pure gluon final state contribution.
The decays $H/A\rightarrow gg$
through a fermion loop
can be found in the literature
\cite{EllGaiNan76,DjoSpiBijZer91,GunHabKanDaw90,Oku82,DjoSpiZer93}.

Using the notation
\beq \EQN{s3}
\ba{ll} \dsp
\beta_f
& \dsp
 = 4m_f^2/M_H^2
\\ \dsp
x_t & \dsp
= \arctan \frac{1}{\sqrt{\beta_t-1}}
\\ \dsp
x_b & \dsp
= \frac{1}{2}\left(
 \pi+i\ln\frac{1+\sqrt{1-\beta_b}}{1-\sqrt{1-\beta_b}}
\right)
\\ \dsp
{\cal M}_f^S
& \dsp = -2\beta_f
 \left[(1-\beta_f)x_f^2 + 1\right]
\\ \dsp
{\cal M}_f^P
& \dsp = \beta_f x_f^2
\ea\eeq
one has
\beq \EQN{s4}
\ba{rl}\dsp
\Delta\Gamma(H\rightarrow gg)
& \dsp  =
 \frac{G_F}{4\sqrt{2}\pi} M_H^3
\left(\frac{\alpha_s}{4\pi}\right)^2
\left\{
2C^S_{tb}\mbox{\rm Re}{\cal M}_t^S
{\cal M}^{S*}_b+
C^S_{bb}|{\cal M}_b^S|^2
\right\}
\\ \dsp
& \dsp
= \frac{3G_F}{4\sqrt{2}\pi}
 M_H  \bar{m}_b^2 \left(\api\right)^2
\\ \dsp
\cdot\Bigg\{ C^S_{tb}\Bigg(
 & \dsp
 \frac{4}{9}+\frac{\pi^2}{9}
-\frac{1}{9}\ln^2\frac{\bar{m}_b^2}{M_H^2}
\\ & \dsp
+\frac{M_H^2}{m_t^2}
\left[ \frac{7}{270}+\frac{7}{1080}\pi^2
-\frac{7}{1080}\ln^2\frac{\bar{m}_b^2}{M_H^2}
\right]
\\ & \dsp
+\frac{\bar{m}_b^2}{M_H^2}
\left[
-\frac{4}{9}\pi^2
- \frac{4}{9}\ln\frac{\bar{m}_b^2}{M_H^2}
+ \frac{4}{9}\ln^2\frac{\bar{m}_b^2}{M_H^2}
\right]
\\ & \dsp
+\frac{\bar{m}_b^2}{m_t^2}
\left[
-\frac{7}{270}\pi^2
- \frac{7}{270}\ln\frac{\bar{m}_b^2}{M_H^2}
+ \frac{7}{270}\ln^2\frac{\bar{m}_b^2}{M_H^2}
\right]
\Bigg)
\\ \dsp
+ C^S_{bb}\Bigg(
 & \dsp
\frac{\bar{m}_b^2}{M_H^2}
\left[
\frac{4}{3}+\frac{2}{3}\pi^2+\frac{1}{12}\pi^4
\right.
\\ & \dsp
\left.
\hphantom{\frac{\bar{m}_b^2}{M_H^2}}
-\left( \frac{2}{3}-\frac{\pi^2}{6}\right)
   \ln^2\frac{\bar{m}^2}{M_H^2}
+\frac{1}{12}
   \ln^4\frac{\bar{m}^2}{M_H^2}
\right]
\Bigg)\Bigg\}
\ea
\eeq

\beq \EQN{s4a}
\ba{rl}\dsp
\Delta\Gamma(A\rightarrow gg)
& \dsp  =
 \frac{G_F}{\sqrt{2}\pi} M_A^3
\left(\frac{\alpha_s}{4\pi}\right)^2
\left\{
2C^P_{tb}\mbox{\rm Re}{\cal M}_t^P
{\cal M}^{P*}_b+
C^P_{bb}|{\cal M}_b^P|^2
\right\}
\\ \dsp
& \dsp
= \frac{3G_F}{4\sqrt{2}\pi}
 M_A \bar{m}_b^2 \left(\api\right)^2
\\ \dsp
\cdot\Bigg\{ C^P_{tb}\Bigg(
 & \dsp
 \frac{1}{6}
-\frac{1}{6}\ln^2\frac{\bar{m}_b^2}{M_A^2}
\\ & \dsp
+\frac{M_A^2}{m_t^2}
\left[ \frac{1}{72}\pi^2
-\frac{1}{72}\ln^2\frac{\bar{m}_b^2}{M_A^2}
\right]
\\ & \dsp
+\frac{\bar{m}_b^2}{M_A^2}
\left[
- \frac{2}{3}\ln\frac{\bar{m}_b^2}{M_A^2}
\right]
\\ & \dsp
+\frac{\bar{m}_b^2}{m_t^2}
\left[
- \frac{1}{18}\ln\frac{\bar{m}_b^2}{M_A^2}
\right]
\Bigg)
\\ \dsp
+ C^P_{bb}\Bigg(
 & \dsp
\frac{\bar{m}_b^2}{M_A^2}
\left[
\frac{4}{3}\pi^4
+\frac{8}{3}\pi^2
   \ln^2\frac{\bar{m}^2}{M_A^2}
+\frac{4}{3}
   \ln^4\frac{\bar{m}^2}{M_A^2}
\right]
\Bigg)\Bigg\}
\ea
\eeq

The singlet contribution to the
 partial width
 of the Higgs bosons into bottom quarks
is therefore given by
\beq \EQN{s5}
\ba{rl}\dsp
\Delta\Gamma^{{\rm {\scriptsize singlet}}}
(H\rightarrow b\bar{b})
& \dsp
= \frac{3G_F}{4\sqrt{2}\pi} M_H \bar{m}_b^2
\left(\api\right)^2
\\ \dsp
\cdot \Bigg\{ C^S_{tb}\Bigg(
& \dsp
\frac{8}{3}-\frac{\pi^2}{9}
-\frac{2}{3}\ln \frac{M_H^2}{m_t^2}
+\frac{1}{9}\ln^2 \frac{\bar{m}_b^2}{M_H^2}
\\ & \dsp
+ \frac{M_H^2}{m_t^2}
\left[
\frac{1381}{24300} - \frac{7}{1080}\pi^2
\right.
\\ & \dsp
\hphantom{+ \frac{M_H^2}{m_t^2}}
\left.
-\frac{41}{1620}\ln\frac{M_H^2}{m_t^2}
+\frac{7}{1080}\ln^2\frac{\bar{m}_b^2}{M_H^2}
\right]
\\ & \dsp
+ \frac{\bar{m}_b^2}{M_H^2}
\left[
 -10 + \frac{4}{9}\pi^2
+4\ln\frac{M_H^2}{m_t^2}
\right.
\\ & \dsp
\hphantom{+ \frac{\bar{m}_b^2}{M_H^2}}
\left.
+\frac{16}{9}\ln\frac{\bar{m}_b^2}{M_H^2}
-\frac{4}{9}\ln^2\frac{\bar{m}_b^2}{M_H^2}
\right]
\\ & \dsp
+ \frac{\bar{m}_b^2}{m_t^2}
\left[
 \frac{713}{2700} + \frac{7}{270}\pi^2
-\frac{7}{270}\ln\frac{M_H^2}{m_t^2}
\right.
\\ & \dsp
\hphantom{+ \frac{\bar{m}_b^2}{M_H^2}}
\left.
+\frac{1}{135}\ln\frac{\bar{m}_b^2}{M_H^2}
-\frac{7}{270}\ln^2\frac{\bar{m}_b^2}{M_H^2}
\right]\Bigg)
\\
\dsp +C^S_{bb} & \dsp
\frac{\bar{m}_b^2}{M_H^2}
\left[
\frac{20}{3}-\frac{2}{3}\pi^2-\frac{1}{12}\pi^4
\right.
\\ & \dsp
\hphantom{\frac{\bar{m}_b^2}{M_H^2}}
\left.
+\left(\frac{2}{3}-\frac{\pi^2}{6}\right)
       \ln^2\frac{\bar{m}_b^2}{M_H^2}
-\frac{1}{12}\ln^4\frac{\bar{m}_b^2}{M_H^2}
\right]
\Bigg\}
\ea
\eeq

\beq \EQN{s5a}
\ba{rl}\dsp
\Delta\Gamma^{{\rm {\scriptsize singlet}}}
(A\rightarrow b\bar{b})
& \dsp
= \frac{3G_F}{4\sqrt{2}\pi} M_A \bar{m}_b^2
\left(\api\right)^2
\\ \dsp
\cdot \Bigg\{ C^P_{tb}\Bigg(
& \dsp
\frac{23}{6}
-\ln \frac{M_A^2}{m_t^2}
+\frac{1}{6}\ln^2 \frac{\bar{m}_b^2}{M_A^2}
\\ & \dsp
+ \frac{M_A^2}{m_t^2}
\left[
\frac{61}{324} - \frac{1}{72}\pi^2
\right.
\\ & \dsp
\hphantom{+ \frac{M_H^2}{m_t^2}}
\left.
-\frac{7}{108}\ln\frac{M_A^2}{m_t^2}
+\frac{1}{72}\ln^2\frac{\bar{m}_b^2}{M_A^2}
\right]
\\ & \dsp
+ \frac{\bar{m}_b^2}{M_A^2}
\left[
 -5
+2\ln\frac{M_A^2}{m_t^2}
-\frac{4}{3}\ln\frac{\bar{m}_b^2}{M_A^2}
\right]
\\ & \dsp
+ \frac{\bar{m}_b^2}{m_t^2}
\left[
 \frac{19}{108}
-\frac{1}{18}\ln\frac{M_A^2}{m_t^2}
-\frac{2}{9}\ln\frac{\bar{m}_b^2}{M_A^2}
\right]\Bigg)
\\
\dsp +C^P_{bb} & \dsp
\frac{\bar{m}_b^2}{M_A^2}
\left[
-\frac{4}{3}\pi^2
-\frac{8}{3}\pi^2
       \ln^2\frac{\bar{m}_b^2}{M_A^2}
-\frac{4}{3}\ln^4\frac{\bar{m}_b^2}{M_A^2}
\right]
\Bigg\}
\ea
\eeq
The distinction between bottom and pure gluon
 final states introduces large logarithms
$\ln^2(m_b^2/M_H^2)$ and
$m_b^2\ln^4(m_b^2/M_H^2)$ in the partial
decay width $\Gamma_{b\bar{b}}$.
Being due to the subtracted
decay into two gluons, their physical origin
might  be traced back to kinematical
configurations which correspond to collinear
gluons with respect to the virtual bottom
quark. This interesting feature deserves
 further analysis in the future.

\section{Nonsinglet Contributions}

Second order QCD corrections to the
scalar and the pseudoscalar
current correlators  were calculated recently
by one of the authors for the case when  the
external momentum
is much larger than  all relevant masses.
The general case of nondiagonal currents
with quarks of different masses was
considered.
A detailed description of this work will
be published in a separate paper \cite{Che95}.
The results can be applied to the Higgs
decay rate in the special case of diagonal
currents.

Besides the three loop diagrams of Figure 2b,
induced contribution from lower order nonsinglet
graphs need to be taken into account, since
they lead to second order corrections after
renormalization of the coupling constant and
the quark mass:
\beq\EQN{n0}
\ba{ll}\dsp
\frac{\alpha_s}{\pi}\Big|_{bare}
& \dsp =
\frac{\alpha_s}{\pi}\left(
1+\frac{\alpha_s}{\pi}
\frac{1}{\epsilon}
\left[-\frac{11}{4}+\frac{1}{6}n_f\right]
\right)
\\ \dsp
m_{bare}
& \dsp
= m\left(
1-\frac{\alpha_s}{\pi}\frac{1}{\epsilon}
\right.
\\ & \dsp
\left.
+\left(\frac{\alpha_s}{\pi}\right)^2
\left[
\frac{1}{\epsilon^2}
\left(\frac{15}{8}-\frac{1}{12}n_f\right)
+\frac{1}{\epsilon}
\left(-\frac{101}{48}+\frac{5}{72}n_f\right)
\right]
\right)
\ea\eeq
For the renormalization of the pseudoscalar
currents the two renormalization constants
$Z^P_5$ and $Z^P_{MS}$ are introduced \cite{Lar92}:
\beq \EQN{n0a}
\ba{ll}
Z^P_5
 & \dsp
= 1-\frac{8}{3}\api + \left(\api\right)^2
\left[
\frac{1}{18}+\frac{1}{27}n_f
\right]
\\
Z^P_{MS}
 & \dsp
= 1-\frac{\alpha_s}{\pi}\frac{1}{\epsilon}
\\ & \dsp
+\left(\frac{\alpha_s}{\pi}\right)^2
\left[
\frac{1}{\epsilon^2}
\left(\frac{15}{8}-\frac{1}{12}n_f\right)
+\frac{1}{\epsilon}
\left(+\frac{25}{16}-\frac{11}{72}n_f\right)
\right]
\ea
\eeq

\noindent

 Ordering the absorptive parts
of the corresponding correlators in massless
and massive contributions
\beq \EQN{n1}
R_{S/P}=R^{(0)}+R^{(m)}_{S/P}
\eeq
one obtains
\beq \EQN{n2}
\ba{ll}\dsp
R^{(0)} =
& \dsp
1
+\left(\api\right)
\left[ \frac{17}{3}-2\ell \right]
\\ & \dsp
+\left(\api\right)^2
\left[
\frac{10801}{144}-\frac{39}{2}\zeta(3)
+\left( -\frac{65}{24}
        +\frac{2}{3}\zeta(3)\right)n'_f
        +\pi^2\left(-\frac{19}{12}
              +\frac{1}{18}n'_f\right)
\right.
\\ & \dsp\left.
\hphantom{+\left(\api\right)^2}
+\ell\left(-\frac{106}{3}+\frac{11}{9}n'_f
       \right)
+\ell^2\left(\frac{19}{4}-\frac{1}{6}n'_f
       \right)
\right]
\ea
\eeq

\beq\EQN{n3}
\ba{ll}\dsp
R^{(m)}_S =
& \dsp
-\frac{6\bar{m}_b^2}{M_H^2}\Bigg(
1+\api
\left[\frac{20}{3}
          -4\ell\right]
\\ & \dsp
\hphantom{-\frac{6m^2}{s}}
+\left(\api\right)^2\Bigg\{
\frac{2383}{24}-\frac{83}{3}\zeta(3)
+\left( -\frac{313}{108}
        +\frac{2}{3}\zeta(3)\right)n'_f
\\ & \dsp
\hphantom{-\frac{6m^2}{s}
+\left(\api\right)^2(}
        +\pi^2\left(-\frac{9}{2}
              +\frac{1}{9}n'_f\right)
+\ell\left(-\frac{371}{6}+\frac{5}{3}n'_f
       \right)
\\ & \dsp
\hphantom{-\frac{6m^2}{s}
+\left(\api\right)^2(}
+\ell^2\left(\frac{27}{2}-\frac{1}{3}n'_f
      \right)
       \Bigg\}\Bigg)
\\ & \dsp
+4\left(\api\right)^2\sum_{f=u,d,s,c,b}
 \frac{m_f^2}{M_H^2}
\ea
\eeq

\beq\EQN{n4}
\ba{ll}\dsp
R^{(m)}_P =
& \dsp
-\frac{2\bar{m}_b^2}{M_A^2}\Bigg(
1+\api
\left[\frac{4}{3}
          -4\ell\right]
\\ & \dsp
\hphantom{-\frac{6m^2}{q^2}}
+\left(\api\right)^2\Bigg\{
\frac{1429}{72}-\frac{83}{3}\zeta(3)
+\left( -\frac{3}{4}
        +\frac{2}{3}\zeta(3)\right)n'_f
\\ & \dsp
\hphantom{-\frac{6m^2}{q^2}
+\left(\api\right)^2(}
        +\pi^2\left(-\frac{9}{2}
              +\frac{1}{9}n'_f\right)
+\ell\left(-\frac{155}{6}+\frac{7}{9}n'_f
       \right)
\\ & \dsp
\hphantom{-\frac{6m^2}{q^2}
+\left(\api\right)^2(}
+\ell^2\left(\frac{27}{2}-\frac{1}{3}n'_f
      \right)
       \Bigg\}\Bigg)
\\ & \dsp
+4\left(\api\right)^2\sum_{f=u,d,s,c,b}
 \frac{m_f^2}{M_A^2}
\ea
\eeq
where $\ell=\ln(M_{H/A}^2/\mu^2)$.
We find agreement with \cite{Sur94a,Sur94b}.

Note that in (\ref{n1}) and (\ref{n2})
$n'_f$ stands for $n_f -1$.  This is because only diagrams
without virtual top lines are included  there. It also implies
that the  renormalization of these contributions  is to be done
by means of (\ref{n0}-\ref{n0a}) with $n_f$ substituted by $n_f'$.

It should be stressed that we are still working in the
full QCD including the top quark and $n_f =6$.
The strong coupling constant
and light quark masses are, thus,  defined with respect to  this theory.
A transition to effective parameters  relevant to  the topless
QCD with five active quark flavours will be  performed
in the next section.

The last terms in eqs. (\ref{n3}),(\ref{n4}) arise
 from double bubble diagrams, where a light quark
is running around a virtual fermion loop.
Although for these diagrams a four fermion
final state ($b\bar{b}f\bar{f}$) is possible,
the assignment to the partial decay rate into
the primarily produced quark flavour
$\Gamma_{b\bar{b}}$ seems to be the natural
choice.

It remains to compute the contribution of
the double bubble diagram
$\Delta\Gamma^{{\rm {\scriptsize DB}}}$
 with a virtual top
quark loop (see Figure 2a).
Again the hard mass procedure can be used
and leads to the result:
\beq\EQN{n5}\ba{rl}\dsp
\Delta\Gamma^{{\rm {\scriptsize DB}}}
(H\rightarrow b\bar{b})
 =
& \dsp
\frac{3G_F}{4\sqrt{2}\pi}M_{H} \bar{m}_b^2
C^{S}_{bb}
\\ \dsp
\cdot \Bigg\{
& \dsp
\frac{89}{216}
+\frac{1}{3}\ln\frac{M_H^2}{m_t^2}\ln\frac{\mu^2}{m_t^2}
-\frac{11}{9}\ln\frac{\mu^2}{m_t^2}
-\frac{1}{6}\ln^2\frac{\mu^2}{m_t^2}
\\ & \dsp
+\frac{M_{H}^2}{m_t^2}
\left[
\frac{107}{675}-\frac{2}{45}
\ln\frac{M_{H}^2}{m_t^2}
\right]
\\ & \dsp
+\frac{\bar{m}_b^2}{M_H^2}
\left[
-\frac{89}{18}
-4\ln\frac{M_H^2}{m_t^2}\ln\frac{\mu^2}{m_t^2}
+10\ln\frac{\mu^2}{m_t^2}
+2\ln^2\frac{\mu^2}{m_t^2}
\right]
\\ & \dsp
+\frac{\bar{m}_b^2}{m_t^2}
\left[
-\frac{116}{75}
+\frac{8}{15}\ln\frac{M_H^2}{m_t^2}
\right]
\Bigg\}
\ea
\eeq

\beq\EQN{n6}\ba{rl}\dsp
\Delta\Gamma^{{\rm {\scriptsize DB}}}
(A\rightarrow b\bar{b})
 =
& \dsp
\frac{3G_F}{4\sqrt{2}\pi}M_{A} \bar{m}_b^2
C^{P}_{bb}
\\ \dsp
\cdot \Bigg\{
& \dsp
\frac{89}{216}
+\frac{1}{3}\ln\frac{M_A^2}{m_t^2}\ln\frac{\mu^2}{m_t^2}
-\frac{11}{9}\ln\frac{\mu^2}{m_t^2}
-\frac{1}{6}\ln^2\frac{\mu^2}{m_t^2}
\\ & \dsp
+\frac{M_{A}^2}{m_t^2}
\left[
\frac{107}{675}-\frac{2}{45}
\ln\frac{M_{A}^2}{m_t^2}
\right]
\\ & \dsp
+\frac{\bar{m}_b^2}{M_A^2}
\left[
-\frac{89}{54}
-\frac{4}{3}\ln\frac{M_A^2}{m_t^2}\ln\frac{\mu^2}{m_t^2}
+\frac{14}{9}\ln\frac{\mu^2}{m_t^2}
+\frac{2}{3}\ln^2\frac{\mu^2}{m_t^2}
\right]
\\ & \dsp
+\frac{\bar{m}_b^2}{m_t^2}
\left[
-\frac{16}{25}
+\frac{4}{15}\ln\frac{M_A^2}{m_t^2}
\right]
\Bigg\}
\ea
\eeq

\section{Transition to the Effective Parameters}

The results obtained above do  not allow a naive  decoupling of the
top quark as they contain the  $\log\mu^2/m_t^2$  terms not suppressed by any
inverse power of $m_t$. From a practical point of view such terms
can hardly be considered as potentially   large  as
$\mu $ is   eventually  set to be equal  to the Higgs boson mass.
Moreover, they may be  summed  up by using the methods of
the effective field theory as it has been recently
demonstrated on the example of top mass
effects in the hadronic decay rate of  the $Z$ boson  \cite{CK4}.

Still it seems to be reasonable to express all our results
in terms of the effective $\alpha_s^{(5)}$ and $m_b^{(5)}$
appearing in the topless QCD with five quark flavours.
The matching equations read
\cite{BerWet82,Ber83,LarRitVer94}:
\beq\EQN{n5a}
\frac{\as^{(6)}(\mu^2)}{\pi}
=
\frac{\as^{(5)}(\mu^2)}{\pi} +
\left(\frac{\as^{(5)}(\mu^2)}{\pi}\right)^2
\frac{1}{6}\ln\frac{\mu^2}{m_t^2}
+ {\cal O}(\as^3)
\eeq

\beq\EQN{n5b}
\bar{m}_b^{(6)}(\mu^2)
=
\bar{m}_b^{(5)}(\mu^2)
\left\{
1-\left(\api\right)^2
\left[
\frac{89}{432} - \frac{5}{36}\ln\frac{\mu^2}{m_t^2}
 + \frac{1}{12}\ln^2\frac{\mu^2}{m_t^2}
\right]\right\}
{},
\eeq

The corresponding changes in our results   can be summarized as follows:
\begin{itemize}
\item
  As the singlet contributions start from order $\alpha_s^2$,  they
      do not change its functional form  within our  accuracy in $\alpha_s$.
\item
Nonsinglet contributions (\ref{n3}) and (\ref{n4})
 are conveniently arranged not to
change their functional form
{\em  if }  new
        $(\alpha_s^{(5)})^2 $  terms ( that are induced by terms of order
        $(\alpha_s^{(6)})^0 $ and $\alpha_s^{(6)} $
        after the replacements (\ref{n0}) and (\ref{n0a}) are done)
        are assigned  to the diagrams with a  virtual
        top loop.
\item
The resulting expressions for the double bubble  contributions
read:
\beq\EQN{eff1}\ba{rl}\dsp
\Delta\Gamma^{{\rm {\scriptsize DB}}}
(H\rightarrow b\bar{b})
 =
& \dsp
\frac{3G_F}{4\sqrt{2}\pi}M_{H} (\bar{m}_b^{(5)})^2
C^{S}_{bb}\left(\api^{(5)}\right)^2
\\ \dsp
\cdot \Bigg\{
& \dsp
\frac{M_{H}^2}{m_t^2}
\left[
\frac{107}{675}-\frac{2}{45}
\ln\frac{M_{H}^2}{m_t^2}
\right]
\\ & \dsp
+\frac{(\bar{m}_b^{(5)})^2}{m_t^2}
\left[
-\frac{116}{75}
+\frac{8}{15}\ln\frac{M_H^2}{m_t^2}
\right]
\Bigg\}
\ea
\eeq

\beq\EQN{eff2}\ba{rl}\dsp
\Delta\Gamma^{{\rm {\scriptsize DB}}}
(A\rightarrow b\bar{b})
 =
& \dsp
\frac{3G_F}{4\sqrt{2}\pi}M_{A} (\bar{m}_b^{(5)})^2
C^{P}_{bb}\left(\api^{(5)}\right)^2
\\ \dsp
\cdot \Bigg\{
& \dsp
\frac{M_{A}^2}{m_t^2}
\left[
\frac{107}{675}-\frac{2}{45}
\ln\frac{M_{A}^2}{m_t^2}
\right]
\\ & \dsp
+\frac{(\bar{m}_b^{(5)})^2}{m_t^2}
\left[
-\frac{16}{25}
+\frac{4}{15}\ln\frac{M_A^2}{m_t^2}
\right]
\Bigg\}
\ea
\eeq
The decoupling of the top quark holds true for
leading as well as subleading bottom mass
corrections. The leading (massless) order
agrees with the expansion of \cite{Kni94}.
\end{itemize}

\section{Discussion}

In this section we
combine the QCD corrections discussed above.
The result
is valid for intermediate Higgs masses (in
particular $M_H<2m_t$),
 since the singlet contribution
was calculated in the heavy top limit.
Everywhere below the notation
 $\alpha_s = \alpha_s^{(5)}(\mu)$,
$\bar{m}_f = \bar{m}_f^{(5)}(\mu)$
is understrood
with $f = u,d,s,c,b$ and  $\mu =M_H$.
Expressing the decay rate in the
following form
\beq \EQN{r1}
\ba{ll}\dsp
\Gamma(H/A\rightarrow b\bar{b})
& \dsp
= \frac{3G_F}{4\sqrt{2}\pi} M_{H/A} \bar{m}_b^2
  C^{S/P}_{bb}
\\ & \dsp
\Bigg\{
1 + \Delta\Gamma_1 \left(\api\right) +
 \left(\api\right)^2
 \left[\Delta\Gamma_2
       +\frac{M_{H/A}^2}{m_t^2}
       \tilde{\Delta\Gamma}_2\right]
\\ & \dsp
+\frac{\bar{m}_b^2}{M_{H/A}^2}\left(
\Delta\Gamma_0^{(m)}
+ \Delta\Gamma_1^{(m)} \left(\api\right)
  +  \left(\api\right)^2
\left[\Delta\Gamma_2^{(m)}
     +\frac{M_{H/A}^2}{m_t^2}\tilde{\Delta\Gamma}_2^{(m)}
\right]
\right)
\\ & \dsp
+{\cal O}\left(\frac{\bar{m}_b^4}{M_{H/A}^4}\right)
\Bigg\}
\ea\eeq
the coefficients for the scalar Higgs partial width
$\Gamma(H\rightarrow b\bar{b})$
read
\beq \EQN{r2}
\ba{ll}
\Delta\Gamma_1\Big|_{H\rightarrow b\bar{b}}
&  \dsp
= \frac{17}{3}
\\
\Delta\Gamma_2\Big|_{H\rightarrow b\bar{b}}
&  \dsp
= 29.147
\\ & \dsp
+\frac{C^S_{tb}}{C^S_{bb}}
\left[
1.57
-\frac{2}{3}\ln\frac{M_H^2}{m_t^2}
+ \frac{1}{9}\ln^2\frac{\bar{m}_b^2}{M_H^2}
\right]
\\
\tilde{\Delta\Gamma}_2\Big|_{H\rightarrow b\bar{b}}
& \dsp
= \frac{107}{675}
-\frac{2}{45}\ln\frac{M_H^2}{m_t^2}
\\  &  \dsp
+\frac{C^S_{tb}}{C^S_{bb}}
\left[
-0.007
-\frac{41}{1620}\ln\frac{M_H^2}{m_t^2}
+ \frac{7}{1080}\ln^2\frac{\bar{m}_b^2}{M_H^2}
\right]
\ea
\eeq

\beq \EQN{r3}
\ba{ll}
\Delta\Gamma_0^{(m)}\Big|_{H\rightarrow b\bar{b}}
&  \dsp
= -6
\\ \dsp
\Delta\Gamma_1^{(m)}\Big|_{H\rightarrow b\bar{b}}
&  \dsp
= -40
\\
\Delta\Gamma_2^{(m)}\Big|_{H\rightarrow b\bar{b}}
&  \dsp
= -107.755
\\ &  \dsp
-0.98\ln^2\frac{\bar{m}_b^2}{M_H^2}
- \frac{1}{12}\ln^4\frac{\bar{m}_b^2}{M_H^2}
 +4\sum_{f=u,d,s,c,b}
         \frac{\bar{m}_f^2}{\bar{m}_b^2}
\\ &  \dsp
+\frac{C^S_{tb}}{C^S_{bb}}
\left[
-5.61
+4\ln\frac{M_H^2}{m_t^2}
+ \frac{16}{9}\ln\frac{\bar{m}_b^2}{M_H^2}
- \frac{4}{9}\ln^2\frac{\bar{m}_b^2}{M_H^2}
\right]
\\
\tilde{\Delta\Gamma}_2^{(m)}\Big|_{H\rightarrow b\bar{b}}
& \dsp
= -\frac{116}{75}
+\frac{8}{45}\ln\frac{M_H^2}{m_t^2}
\\ &  \dsp
+\frac{C^S_{tb}}{C^S_{bb}}
\left[
0.52
-\frac{7}{270}\ln\frac{M_H^2}{m_t^2}
+ \frac{1}{135}\ln\frac{\bar{m}_b^2}{M_H^2}
- \frac{7}{270}\ln^2\frac{\bar{m}_b^2}{M_H^2}
\right]
\ea
\eeq
For the decay rate
$\Gamma(A\rightarrow b\bar{b})$
of the pseudoscalar Higgs  boson one has
($\mu^2=M_A^2$)
\beq \EQN{r4}
\ba{ll}
\Delta\Gamma_1\Big|_{A\rightarrow b\bar{b}}
&  \dsp
= \frac{17}{3}
\\
\Delta\Gamma_2\Big|_{A\rightarrow b\bar{b}}
&  \dsp
= 29.147
\\ &  \dsp
+\frac{C^P_{tb}}{C^P_{bb}}
\left[
\frac{23}{6}
-\ln\frac{M_A^2}{m_t^2}
+ \frac{1}{6}\ln^2\frac{\bar{m}_b^2}{M_A^2}
\right]
\\
\tilde{\Delta\Gamma}_2\Big|_{A\rightarrow b\bar{b}}
& \dsp
= \frac{107}{675}
-\frac{2}{45}\ln\frac{M_A^2}{m_t^2}
\\  &  \dsp
+\frac{C^P_{tb}}{C^P_{bb}}
\left[
0.051
-\frac{7}{108}\ln\frac{M_A^2}{m_t^2}
+ \frac{1}{72}\ln^2\frac{\bar{m}_b^2}{M_A^2}
\right]
\ea
\eeq

\beq \EQN{r5}
\ba{ll}
\Delta\Gamma_0^{(m)}\Big|_{A\rightarrow b\bar{b}}
&  \dsp
= -2
\\ \dsp
\Delta\Gamma_1^{(m)}\Big|_{A\rightarrow b\bar{b}}
&  \dsp
= -\frac{8}{3}
\\
\Delta\Gamma_2^{(m)}\Big|_{A\rightarrow b\bar{b}}
&  \dsp
= 91.006
\\ &  \dsp
-26.32\ln^2\frac{\bar{m}_b^2}{M_A^2}
- \frac{4}{3}\ln^4\frac{\bar{m}_b^2}{M_A^2}
 +4\sum_{f=u,d,s,c,b}
         \frac{\bar{m}_f^2}{\bar{m}_b^2}
\\ &  \dsp
+\frac{C^P_{tb}}{C^P_{bb}}
\left[
-5
+2\ln\frac{M_A^2}{m_t^2}
- \frac{4}{3}\ln\frac{\bar{m}_b^2}{M_A^2}
\right]
\\
\tilde{\Delta\Gamma}_2^{(m)}\Big|_{A\rightarrow b\bar{b}}
& \dsp
= -\frac{16}{25}
+\frac{4}{15}\ln\frac{M_A^2}{m_t^2}
\\ &  \dsp
+\frac{C^P_{tb}}{C^P_{bb}}
\left[
\frac{19}{108}
-\frac{1}{18}\ln\frac{M_A^2}{m_t^2}
- \frac{2}{9}\ln\frac{\bar{m}_b^2}{M_A^2}
\right]
\ea
\eeq
The formula for the decay rate of the
standard scalar Higgs boson in the
minimal SM may be obtained by
setting $C^S_{bb}=C^S_{bb}=1$.
{}From eqs. (\ref{r2}),(\ref{r3})
one derives
\beq \EQN{r6}
\ba{ll}
\Delta\Gamma_1
\Big|_{H\rightarrow b\bar{b}}^{{\rm SM}}
&  \dsp
= \frac{17}{3}
\\
\Delta\Gamma_2
\Big|_{H\rightarrow b\bar{b}}^{{\rm SM}}
&  \dsp
= 30.717
-\frac{2}{3}\ln\frac{M_H^2}{m_t^2}
+ \frac{1}{9}\ln^2\frac{\bar{m}_b^2}{M_H^2}
\\
\tilde{\Delta\Gamma}_2
\Big|_{H\rightarrow b\bar{b}}^{{\rm SM}}
& \dsp
= 0.15
-\frac{113}{1620}\ln\frac{M_H^2}{m_t^2}
+ \frac{7}{1080}\ln^2\frac{\bar{m}_b^2}{M_H^2}
\ea
\eeq

\beq \EQN{r7}
\ba{ll}
\Delta\Gamma_0^{(m)}
\Big|_{H\rightarrow b\bar{b}}^{{\rm SM}}
&  \dsp
= -6
\\ \dsp
\Delta\Gamma_1^{(m)}
\Big|_{H\rightarrow b\bar{b}}^{{\rm SM}}
&  \dsp
= -40
\\
\Delta\Gamma_2^{(m)}
\Big|_{H\rightarrow b\bar{b}}^{{\rm SM}}
&  \dsp
= -113.369
+4\ln\frac{M_H^2}{m_t^2}
\\ &  \dsp
+ \frac{16}{9}\ln\frac{\bar{m}_b^2}{M_H^2}
- 1.42\ln^2\frac{\bar{m}_b^2}{M_H^2}
- \frac{1}{12}\ln^4\frac{\bar{m}_b^2}{M_H^2}
\\ &  \dsp
 +4\sum_{f=u,d,s,c,b}
         \frac{\bar{m}_f^2}{\bar{m}_b^2}
\\
\tilde{\Delta\Gamma}_2^{(m)}
\Big|_{H\rightarrow b\bar{b}}^{{\rm SM}}
& \dsp
= -1.027
+\frac{137}{270}\ln\frac{M_H^2}{m_t^2}
+ \frac{1}{135}\ln\frac{\bar{m}_b^2}{M_H^2}
- \frac{7}{270}\ln^2\frac{\bar{m}_b^2}{M_H^2}
\ea
\eeq
For our numerical discussion we use as input
 parameters a top mass of $m_t=176$ GeV and a
bottom pole mass of $m_b=4.7$ GeV.
The latter translates into the running mass
$\bar{m}_b(M_H^2)=2.84/2.75/2.69$ GeV for Higgs
masses of $M_H=70/100/130$ GeV.
All other quarks are assumed to be massless. Based on
$\Lambda_{QCD}=233$ MeV one arrives at the following
values for the strong coupling constant:
$\as(M_H^2)=0.125/0.118/0.114$ corresponding
to the three differerent values of $M_H$.
The influence of the top quark
induced contribution
on the
second order coefficient is significant.
 $\Delta\Gamma_2$
is shifted
 to $(36.60/37.39/38.12)(\alpha_s/\pi)^2$
and is mainly due to the double triangle
contribution  with its mass logarithms.
 The quadratic bottom mass corrections prove to
be rather small. They add a contribution
of $(-0.54/-0.32/-0.22)(\as/\pi)^2$ to the second order
result.
Finally we reproduce the expression for the
total hadronic width
 of the Higgs boson
in the minimal SM:
\beq \EQN{r8}
\ba{ll}\dsp
\Gamma^{{\rm SM}}(H\rightarrow {\rm hadrons})
& \dsp
= \frac{3G_F}{4\sqrt{2}\pi} M_{H}
\sum_{f=u,d,s,c,b}\bar{m}_f^2
\\ & \dsp
\Bigg\{
1 + \frac{17}{3} \left(\api\right)
\\ & \dsp
+ \left(\api\right)^2
 \left[32.26-\frac{2}{3}\ln\frac{M_{H}^2}{m_t^2}
\right]
\\ & \dsp
\hphantom{+ \left(\api\right)^2}
+\frac{M_H^2}{m_t^2}
\left[
\frac{5863}{24300}
-\frac{113}{1620}\ln\frac{M_{H}^2}{m_t^2}
\right]
\\ & \dsp
+\frac{\bar{m}_f^2}{M_H^2}
\left(
-6 - 40 \left(\api\right)
\right.
\\ & \dsp
\hphantom{\frac{\bar{m}_f^2}{M_H^2}}
+ \left(\api\right)^2
 \left[
-109.72+4\ln\frac{M_{H}^2}{m_t^2}
+\frac{4}{3}\ln\frac{\bar{m}_f^2}{M_H^2}
\right]
\\ & \dsp
\hphantom{+\left(\api\right)^2\frac{\bar{m}_f^2}{M_H^2}}
\left.
+\frac{M_H^2}{m_t^2}
\left[
 -1.28
+\frac{137}{270}\ln\frac{M_H^2}{m_t^2}
- \frac{1}{54}\ln\frac{\bar{m}_f^2}{M_H^2}
\right]
\right)
\\ & \dsp
+\sum_{f'=u,d,s,c,b}
12\frac{\bar{m}_{f'}^2}{M_H^2}
\left(\api\right)^2
\Bigg\}
\ea\eeq
Contrary to the partial rate into bottom quarks the
hadronic width is practically not
affected by large logarithms $\ln(\bar{m}_f^2/M_H^2)$,
since they vanish in the sum of bottom and gluon final states.
Their  absence is reflected in the numerical numbers for
the second order coefficients. For the massless
corrections they read
 $(33.55/33.12/32.82)/(\alpha_s/\pi)^2$. The
mass corrections are again small. They amount to
($-0.19/-0.085/-0.048)(\as^2/\pi)^2$.

\vspace{2ex}
\noindent
{\large {\bf Acknowledgments}}

\noindent
We are indebted to  J.H.~K\"uhn for helpful discussions.
K.~Chetyrkin
thanks  the Universit\"at Karlsruhe for a  guest
professorship and  Institute of Theoretical Particle
Physics for warm hospitality.
The authors thank the
Deutsche Forschungsgemeinschaft for financial support
(grants no. Ku 502/6-1  and Kw 8/1-1 ). Partial support by US DOE under
Contract DE-AC03-76SF00098 is gratefully acknowledged.

\pagebreak

\end{document}